\documentclass[conference]{IEEEtran}
\IEEEoverridecommandlockouts

\usepackage[letterpaper, left=0.625in, right=0.625in, top=0.7in, bottom=1in]{geometry}
\usepackage{lipsum}  

\usepackage{booktabs}
\usepackage{url}
\usepackage{siunitx}
\usepackage[table]{xcolor} 
\usepackage{tabularx}      
\usepackage{cite}
\usepackage{amsmath,amssymb,amsfonts}
\usepackage{algorithmic}
\usepackage{subcaption}
\usepackage{graphicx}
\usepackage{subcaption}
\usepackage{textcomp}
\usepackage{xcolor}
\usepackage{soul}
\usepackage{float}
\usepackage{etoolbox}
\makeatletter
\patchcmd{\@float}{\@fpsadddefault}{\@fpsadddefault\vspace{-0.8\baselineskip}}{}{}
\makeatother
\usepackage{multirow}
\def\BibTeX{{\rm B\kern-.05em{\sc i\kern-.025em b}\kern-.08em
    T\kern-.1667em\lower.7ex\hbox{E}\kern-.125emX}}
\begin{document}

\title{Characterization of FR3 Cellular Vehicle-to-Base Station Links in HighRise Urban Scenarios
\\
}
\author{
Fahimeh Aghaei$^{*}$, Mehdi Monemi$^{*}$, Mehdi Rasti$^{*}$, Murat Uysal$^{\dagger,\ddagger}$\\[1ex]
\small $^{*}$\textit{Centre for Wireless Communications (CWC),  University of Oulu, Oulu, Finland}\\
\small $^{\dagger}$\textit{Division of Engineering, New York University Abu Dhabi (NYUAD), Abu Dhabi 129188, UAE}\\
\small $^{\ddagger}$\textit{Research Institute NYUAD Wireless Center, Abu Dhabi 129188, UAE}\\

\small fahimeh.aghaei@oulu.fi, mehdi.monemi@oulu.fi,  mehdi.rasti@oulu.fi, murat.uysal@nyu.edu
\thanks{This paper is supported by the Finnish Ministry of Education and Culture through the Intelligent Work Machines Doctoral Education Pilot Program (IWM VN/3137/2024-OKM-4); the Research Council of Finland (former Academy of Finland) via 6G Flagship (Grant No. 369116) and Profi6 Project (Grant No. 336449); and the Business Finland via the 6GBridge - Local 6G project (Grant No. 8002/31/2022) and DigiPave project with (Grant No. 3992/31/2023). The work of M. Uysal is supported by Tamkeen under the Research Institute NYUAD grant CG017.}}

\maketitle
\begin{abstract}
Driven by the escalating demand for wireless capacity and advancements in 6G research, the new Frequency Range 3 (FR3) referred to upper mid-band (7.125-24.25 GHz) has emerged as a highly compelling spectrum candidate. This range offers a trade-off exploiting the high bandwidth capabilities of millimeter wave frequencies and the superior propagation characteristics of sub-6 GHz bands. As such, the upper mid-band presents an opportunity to enhance both coverage and capacity particularly in the context of 6G and Cellular Vehicle-to-Base Station (C-V2B). Crucially, realizing this potential requires overcoming technical challenges through accurate and realistic channel modeling, especially in dense, high-rise urban environments.
To address this, we employ a ray-tracing tool to analyze downlink propagation characteristics, enabling detailed channel modeling for reliable C-V2B communication. Our analysis evaluates the signal-to-noise ratio (SNR) and signal-to-interference-plus-noise ratio (SINR) across sub-6 GHz, FR3, and mmWave bands using antenna array configurations designed for high-rise urban areas. Results show that, under equal aperture sizes across frequencies, FR3 achieves superior SNR compared to mmWave in interference-free conditions. Moreover, under the full-interference case, FR3 yields higher SINR for cell-edge User Equipment (UEs). This indicates that the increased array gain at mmWave cannot fully compensate for the severe path loss experienced by cell-edge UEs.
\end{abstract}
\section{Introduction}
Ongoing innovations in wireless communication technologies are catalyzing the evolution of the sixth-generation (6G) networks to support key use cases such as ultra-reliable low-latency communication (URLLC), massive machine-type communications (mMTC), and high-capacity services in dynamic environments. This evolution is further fueled by the exponential growth in data demand driven by emerging services such as fixed wireless access (FWA), extended reality (XR), metaverse applications, and mobile gaming. Recent analyses indicate that 5G users consume up to 2.7 times more data than 4G users \cite{Ref1}, while FWA and metaverse services can demand 23 and 4 times more data compared to conventional mobile users, respectively. To meet these demands, the Frequency Range 3 (FR3, 7.125–24.25 GHz) has emerged as a promising 6G candidate\cite{RefFR3} balancing the large bandwidths available at millimeter-wave (mmWave) frequencies (i.e., FR2) and the superior propagation characteristics of sub-6 GHz bands (i.e., FR1) \cite{SurveyFr3}. 

Urban performance analysis of 6G networks evaluates Key Performance Indicators (KPIs) such as signal quality and coverage in areas with varying building densities and heights \cite{Ref23}. HighRise Urban environments are particularly challenging due to tall, irregular structures that intensify multipath propagation, scattering, diffraction, and attenuation. Frequent transitions between Line-of-Sight (LoS) and Non-Line-of-Sight (NLoS) links further degrade signal stability, making accurate channel modeling essential for reliable 6G evaluation.

Accurate channel models are crucial for designing and assessing next-generation wireless systems in dense cities. The 3rd Generation Partnership Project (3GPP) TR 38.901 report provides standardized stochastic models across sub-6 GHz, mmWave, and upper mid-band frequencies, capturing effects like path loss and shadowing \cite{Ref3}. However, while useful for KPI analysis, these models lack the spatial and temporal resolution needed to fully optimize advanced technologies such as massive Multiple-Input and Multiple-Output (MIMO) and beamforming for both vehicular and non-vehicular communications \cite{monemi2025higher, RefStochasticRaytracing}.

Recent studies have explored wireless channel behavior across different frequency ranges and deployment conditions to support emerging 6G applications. The authors in \cite{Refmatlab} investigated air-to-ground channel characteristics at 2.4 GHz for aerial Base Stations (BSs) using Ray Tracing (RT), revealing environment-dependent path loss exponents and shadow fading parameters across SubUrban, Urban, and Urban High-Rise scenarios. Complementing this, \cite{RappFr3} conducted comprehensive measurement campaigns at 6.75 GHz (FR1C) and 16.95 GHz (FR3) in urban microcell environments. Analysis of the path loss reveals lower directional and omnidirectional path loss exponents in upper mid-band compared to mmWave and sub-THz frequencies. Additionally, a decreasing trend in Root Mean Square (RMS) delay spread and angular spread by frequency growth was observed. Similarly, \cite{MainRef} emphasized the importance of upper mid-band spectrum for achieving a trade-off between coverage and capacity, motivating the need for accurate channel models to guide system design under realistic propagation conditions.

In this paper, we present a comprehensive FR3 channel characterization in a HighRise Urban environment compared with sub-6 GHz and mmWave bands. Using RT techniques, we focuse on Cellular Vehicle-to-Base Station (C-V2B) downlink communication scenarios, where BSs transmit to vehicles acting as User Equipments (UEs)\footnote{Considering UEs with identical antenna gains, the performance in vehicular networks typically falls between that of fixed-mounted devices (e.g., fixed sensors) which benefit from stable placement and the absence of metallic body of the vehicle, and that of handheld UEs (e.g., pedestrians holding cellphones), which experience higher signal variability due to hand grip attenuation. 
}.
Motivated by \cite{Ref8, Ref9} employing RT for mmWave and vehicular studies, we adopt this approach with detailed 3D CAD models of vehicles and the different urban layouts. MIMO technology is integrated to assess beamforming impacts on signal quality, data rate, and coverage. Simulations are conducted using Remcom’s Wireless InSite \cite{Ref6} to evaluate signal-to-noise ratio (SNR) and signal-to-interference-plus-noise ratio (SINR) under different interference scenarios. The main contributions of this work are:

\begin{itemize}
    \item To enable the integration of the FR3 band into future 6G networks, we conduct a comprehensive C-V2B performance analysis of FR3 frequencies (8.2 GHz and 15 GHz) in comparison with sub-6 GHz (4.6 GHz) and mmWave (28 GHz) bands using a RT tool. The study considers both interference-free and full-interference cases employing a static blockage model resulted from buildings based on the statistical ITU model in HighRise Urban environments. Furthermore, a realistic 3D CAD model of Dubai downtown is incorporated and validated against the ITU model. 
    To ensure a fair comparison, we assume an equal aperture size across all frequencies. {\it The comparison of the SINR between the FR3 and mmWave shows that for {\bf cell-edge} UEs, where the average SINR is typically the lowest, FR3 provides a higher performance. Therefore,  although mmWave offers a higher array gain due to the incorporation of larger number antenna elements, this gain cannot fully offset the severe path loss at mmWave frequencies.
    }
    
     \item While an increase in BS density evidently enhances the coverage probability in {\it interference-free} scenario across all frequencies, our extensive simulations demonstrate that this trend does not hold under {\it full-interference} scenario.  
     For example, given a 10 dB coverage probability SINR threshold \cite{sinr10}, the SINR performance at higher frequencies (15 GHz and 28 GHz) shows a monotonically increasing behavior with rising BS density. Conversely, the performance at lower frequencies (8.2 GHz and 4.6 GHz) peaks at an optimal density of 17 BS/km$^{2}$ before dropping.  
\end{itemize}

The organization of the paper is as follows: Section II describes the city layout and scenarios, followed by the channel modeling methodology in Section III. Statistical evaluation of downlink C-V2B performance is presented in Section IV, and Section V offers concluding remarks.

\section{City Layout and Scenarios}
\begin{figure}[t!]
    \centering
    \begin{subfigure}[t]{0.44\linewidth}
        \centering
        \includegraphics[width=\linewidth, clip, trim=10 2 0 40]{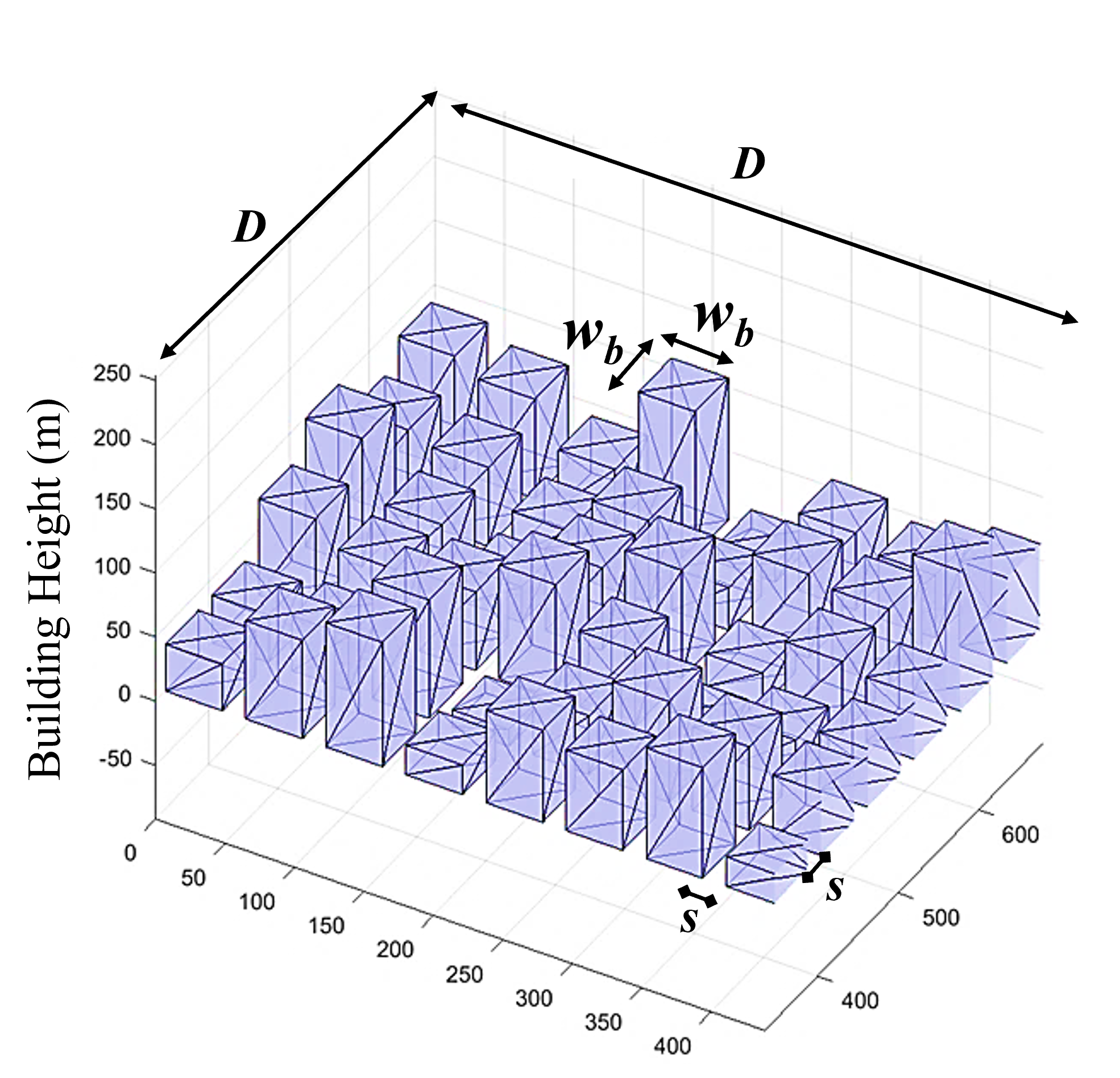}
        \caption*{(a)}
    \end{subfigure}
    \hfill
    \hspace{-5.6em}
    \begin{subfigure}[t]{0.49\linewidth}
        \centering
        \includegraphics[width=\linewidth, clip, trim=86 2 0 10]{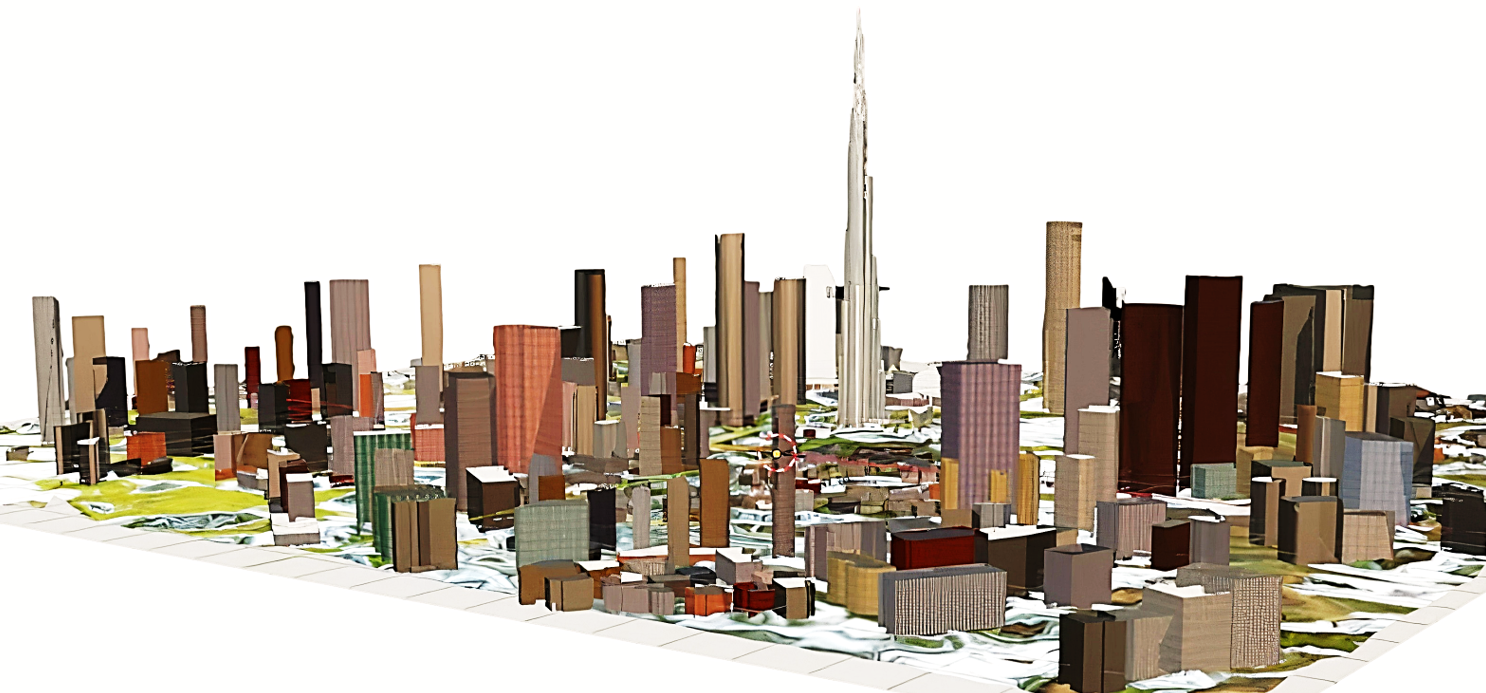}
        \caption*{(b)}
    \end{subfigure}
    \caption{\textcolor{black}{Generated city model: (a) Based on the ITU statistical model, (b) 3D CAD model of the Dubai downtown area.}}
    \label{fig:Math}
\end{figure}

Accurate channel modeling in urban environments requires a clearly well-defined set of parameters, especially those describing the layout and characteristics of urban buildings. To support this, the International Telecommunication Union (ITU-R) recommends a standardized urban model based on three key parameters: $\alpha_0$, representing the ratio of the built-up land area to the total land area; $\beta_0$, the average number of building density measured in buildings per square kilometer; and $\gamma_0$, a scale parameter representing the distribution of building heights using a Rayleigh probability density function\cite{RefITUTerr}:
\begin{equation}
P(h_\mathrm{b}) = \frac{h_\mathrm{b}}{\gamma_0}\exp \left(\frac{ - h_\mathrm{b}^2}{2\gamma_0^2}\right),
\label{EQ:1}
\end{equation} 
where ${h_\mathrm{b}}$ is the height of the building. Together, these parameters effectively capture the essential geometrical characteristics of urban environments, enabling more reliable propagation modeling. The models are particularly valuable in areas with high building density and tall structures, such as HighRise Urban environments that feature a mix of mid- and high-rise buildings. Although building heights are randomly generated according to (1), the building width and spacing are kept constant. The corresponding city-layout parameters such as the building width (${w_\mathrm{b}}$) in meters, the street width (${s}$) in meters, the network area (${D}$) in kilometers for a square patch, and the number of buildings within the patch (${N_\mathrm{b}}$) also are linked to the ITU statistical parameters as follows: $\alpha_0 = \frac{w_\mathrm{b}^{2} \cdot N_\mathrm{b}}{(1000D)^{2}}$, $\beta_0 = \frac{N_\mathrm{b}}{D^2}$, $w_\mathrm{b} = 1000\sqrt{\frac{\alpha_0}{\beta_0}}$, $D = \frac{s + w_\mathrm{b}}{1000}\sqrt{N_\mathrm{b}}$, and $s = \frac{1000}{\sqrt{\beta_0}} - w_\mathrm{b}$. For clarity, Fig. \ref{fig:Math}(a) illustrates an example generated area and corresponding parameters within a limited region. Using this approach, the realization of a hypothetical city layout has been sampled from the ITU statistical model and then imported to the RT. The ITU statistical parameters are set as $\alpha_{0} = 0.5$, $\beta_{0} = 300$, and $\gamma_{0} = 50$. 
These parameters correspond to a network area of $D = 1.2\,\text{km}^2$, 
containing approximately ${N_\mathrm{b}} = 432$ buildings. The building width and street width are kept constant at ${w_\mathrm{b}} = 40.82\,\text{m}$ and $s = 16.91\,\text{m}$, respectively. To validate our findings, we employed a 3D CAD model of Dubai downtown representing a HighRise Urban area generated by Blender-Open Street Map (OSM) software. This tool integrates urban geometry with real-world features and material properties, enabling a realistic representation of the environment, as illustrated in Fig. \ref{fig:Math}(b). 
To assess C-V2B performance in realistic settings with static blockages from buildings, the system behavior is analyzed under varying two interference scenarios as follows:
\begin{itemize}
\item Scenario 1 ({\it interference-free}): Each UE is scheduled on orthogonal resources, experiencing no inter-cell interference. The received signal quality depends only on thermal noise and propagation characteristics.
\item Scenario 2 ({\it full-Interference}): All BSs reuse the same frequency resources, causing each UE to experience interference from all non-serving BSs, representing the maximum possible interference.
\end{itemize}
\section{Channel Modeling Methodology}
 
We consider a MIMO downlink system with $N_t$ transmit antennas per sector at each BS and $N_r$ receive antennas at each UE. A Uniform Rectangular Array (URA) is employed at the BS, while a Uniform Linear Array (ULA) is applied at the receiver. MIMO channel models are generally classified as either narrowband or wideband, depending on system bandwidth. Narrowband models assume a frequency-flat response across the whole bandwidth, whereas wideband models account for frequency-selective fading induced by multipath propagation. The choice between a narrowband or wideband channel model depends on the channel's coherence bandwidth $B_\mathrm{c}$. In various urban environments, multipath propagation results in RMS delay spread $\tau_{\text{rms}}$ ranging from approximately 74.5 to 490 ns within the 2.1-28.5 GHz frequency range, as reported in ITU-R P.1411-10 (Table 11)~\cite{ITURef}. Noting that if $B_\mathrm{c}$ is defined as the bandwidth over which the frequency correlation function is above 0.9, then $B_\mathrm{c}$ is approximately $B_\mathrm{c} \approx  1/(5\tau_{\text{rms}})$. 
Therefore, $B_\mathrm{c}$ in our corresponding frequency range varies from order of tens of kHz to a few tens of MHz. Based on 3GPP specifications, practical bandwidths in these frequencies commonly range from 20 MHz to 400 MHz. Since these bandwidths significantly exceed $B_\mathrm{c}$, the channel is modeled as frequency-selective, necessitating the adoption of a wideband MIMO channel model. 

Consider a wideband MIMO system, where the corresponding baseband input–output relationship is expressed as:
\begin{equation}
\mathbf{y}(t) = \int \mathbf{H}(\tau)\,\mathbf{x}(t-\tau)\,d\tau + \mathbf{n}(t),
\end{equation}
where $\mathbf{x}(t)$ denotes the transmitted signal, 
$\mathbf{y}(t)$ is the received signal, $\mathbf{n}(t)$ represents additive white Gaussian noise (AWGN), and and $\mathbf{H}(\tau)$ is the
$N_\mathrm{r} \times N_\mathrm{t}$ wideband MIMO channel impulse response matrix.

The channel matrix $\mathbf{H}(\tau)$, which characterizes the
complex channel gains between the transmit and receive antenna
elements, is expressed as:
\begin{equation}
\mathbf{H}(\tau) =
\begin{bmatrix}
h_{1,1}(\tau) & \cdots & h_{1,N_\mathrm{t}}(\tau) \\
\vdots       & \ddots & \vdots        \\
h_{N_\mathrm{r},1}(\tau) & \cdots & h_{N_\mathrm{r},N_\mathrm{t}}(\tau)
\end{bmatrix}_{N_\mathrm{r} \times N_\mathrm{t}},
\end{equation} 
where each channel coefficient $h_{mn}(\tau)$ models the link from the
$n$-th transmit antenna to the $m$-th receive antenna and is given by:
\vspace{-0.5em}
\begin{equation}
h_{m,n}(\tau) = \sum_{i=1}^{N_\mathrm{p}} A^{(i)}_{m,n}\,
e^{j \psi^{(i)}_{m,n}}\, \delta\!\big(\tau - \tau^{(i)}_{m,n}\big),
\end{equation}
where $\delta(\cdot)$ denotes the Dirac delta function, $N_\mathrm{p}$ is the number
of multipath components between transmit antenna $n$ and receive
antenna $m$, and $A^{(i)}_{m,n}$, $\psi^{(i)}_{m,n}$, and $\tau^{(i)}_{m,n}$
represent the amplitude, phase, and delay of the $i$-th path on the
$m$–$n$ link, respectively.
\vspace{-0.5em}
\section{Statistical Evaluation of Downlink C-V2B Performance}
\label{sec:IV}
In this section, we present a statistical evaluation of downlink C-V2B performance in HighRise Urban areas under both {\it interference-free} and {\it full-interference} scenarios considering static blockage. The analysis focuses on CDFs of SNR, and SINR across different frequencies. The Maximum Ratio Transmission (MRT) is adopted for transmit beamforming in a single-stream MIMO configuration. 
At the receiver side, diversity is modeled using Maximal Ratio Combining (MRC). 
 Each elements of the beamforming vector is normalized so that the sum of its squared magnitudes equals the number of receive antennas $N_{\mathrm{r}}$, ensuring fair comparison across combining schemes. In the following, we present the formulations used to evaluate the leverage of the RT in a wideband MIMO communication setup. 
In the presence of full interference from all neighboring BSs, i,e., under {\it full-interference} scenario, the system performance is quantified by the SINR, expressed as \cite{Ref6}:
\vspace{-0.5em}
\begin{equation}
\mathrm{SINR} \;=\; \frac{P_{\mathrm{t}}}{N_\mathrm{t}\,(\sigma_n^2+P_{\text{I,avg}})}\,
\sum_{m=1}^{N_r}\sum_{n=1}^{N_t} \bigl| \mathbf{H}_{m,n} \bigr|^2,
\end{equation}
\noindent where $P_{\mathrm{t}}$ is the total transmit power, $\sigma_n^2$ =$N_{0}$$B$ denotes the noise variance, $N_{0}$ is thermal noise spectral density ($10^{-21} \rm{A}^{2}/\rm{Hz}$), $B$ is the channel bandwidth, and $P_\mathrm{I,avg}$ represents the average interference power from $N_\mathrm{I}$ interfering BSs, given by 
\begin{equation}
P_{\text{I,avg}} \;=\; 
\sum_{j=1}^{N_{\mathrm{I}}} \frac{P_{\mathrm{t},j}}{N_{\mathrm{t},j}} 
\left[ \frac{1}{N_\mathrm{r}} \sum_{u=1}^{N_{\mathrm{t},j}} \sum_{m=1}^{N_\mathrm{r}} |\mathbf{H}_{j,u,m}|^2 \right], 
\end{equation}
\noindent with $P_{\mathrm{t},{j}}$ denotes the transmit power of the $j$-th interfering BS, $N_{\mathrm{t},{j}}$ is the number of the $j$-th interfering BS  elements, and $\mathbf{H}_{j,u,m}$ is the channel coefficient between the $u$-th element of $j$-th BS and the $m$-th element of the UE. Note that in the absence of average interference power, i.e., when $P_\mathrm{I,avg}$\,=\,0, the SINR simplifies to SNR. 

\begin{figure}[t!]
\centering
\includegraphics[clip,trim=2 1 6 3, scale=0.27]{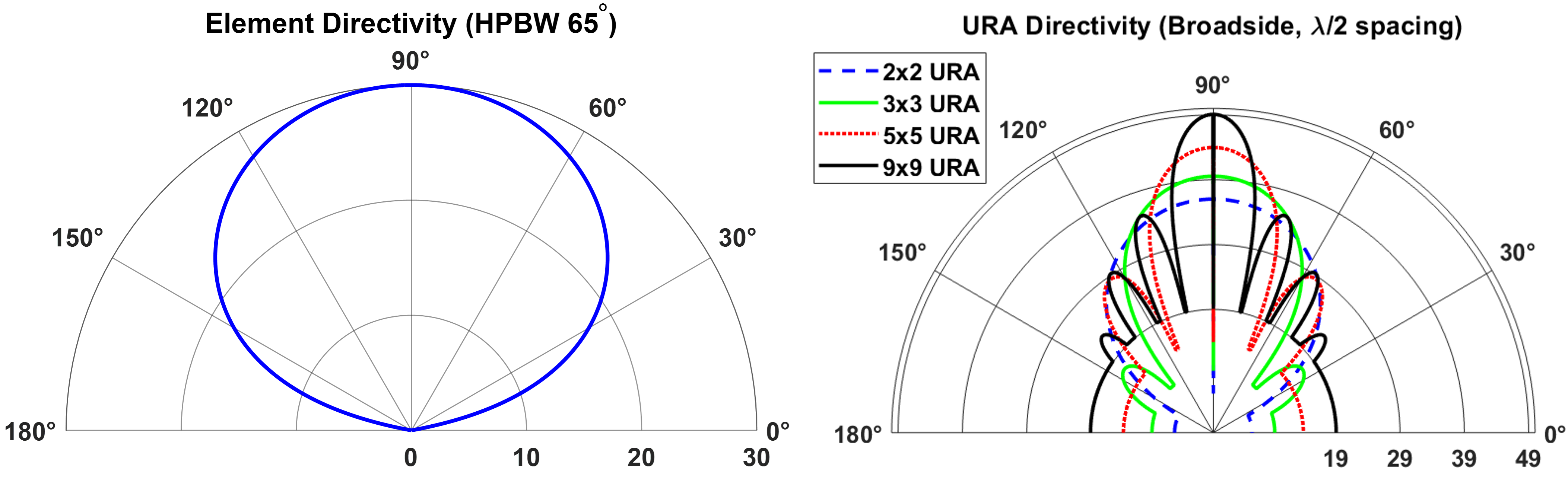}
\caption{Directivity patterns for single antenna element and antenna arrays of BS configurations with 2×2, 3×3, 5×5, and 9×9 URA.}
\label{fig:AntPat}
 \end{figure}

\begin{figure}[t!]
\centering
\includegraphics[clip,trim=7 15 0 5, scale=0.22]{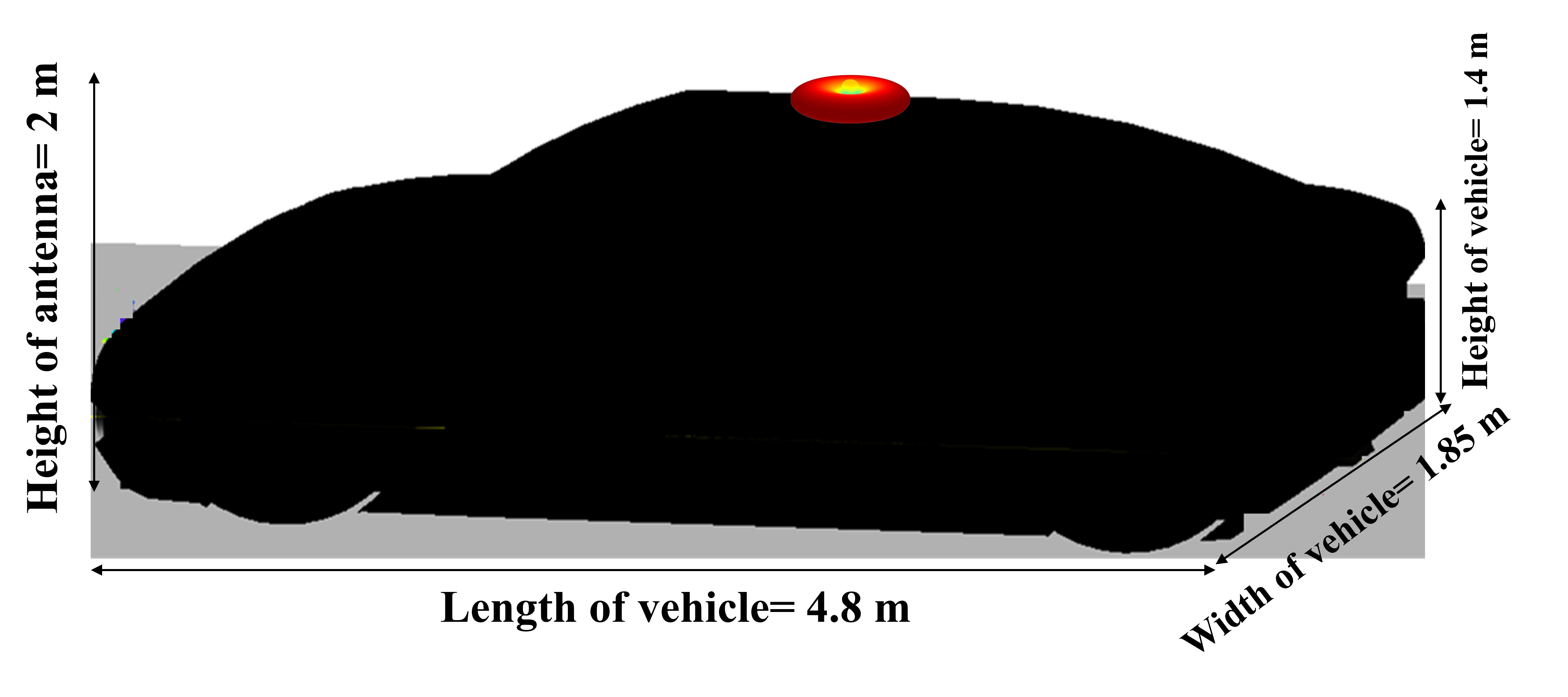}
\caption{\textcolor{black}{Vehicle CAD model with antenna placement.}}
\label{fig:VCAD}
 \end{figure}

 In all simulations the serving BS for each UE is assigned as the one providing the highest signal quality. 
The evaluations are conducted using both ITU statistical parameters, and the 3D CAD model of the Dubai downtown area corresponding to a realistic HighRise Urban city layout illustrated in Fig. \ref{fig:Math}(a) and (b), respectively. The simulator is configured to allow up to 6 reflections, 1 diffraction, and 1 transmission per path with a maximum of 25 paths per link, while capturing only those with power levels above~-250 dBm. Table \ref{tab:table2} lists the conductivity (S/m) and relative permittivity (denoted as $\sigma$ and $\varepsilon_r$) of various materials including concrete, glass, wood, brick, and dry earth ground across different frequencies \cite{Ref10}. All simulations are carried out for 1.2$\times$1.2 km\textsuperscript{2} network area through the Wireless Insite. Within the network area, 17 BSs are deployed in a hexagonal grid pattern with an Inter Site Distance (ISD) of 350 m\cite{Ref3}. The rooftop-mounted antennas have a Half-Power BeamWidth (HPBW) of 65° and a maximum element gain of 30 dBi with a downtilt of –12°. The URA dimensions are scaled with frequency to keep a same transmitter aperture size ensuring a fair performance comparison \cite{MainRef}. Considering a half-wavelength inter-element spacing, the number of BS antenna elements corresponding to each frequency is presented in Table \ref{tab:table3}, which aligns with \cite{MainRef}. Fig. \ref{fig:AntPat} illustrates the element-wise and array-wise directivity patterns for the BS, where the aperture directivity is presented for various frequencies. The allocated bandwidth also increases with the carrier frequency, reflecting standardized spectrum allocation policies in wireless systems \cite{Ref3}. For accurate RT simulations, 370 UEs are uniformly distributed across the area, with results averaged over 10 city deployments. Fig. \ref{fig:VCAD} shows the vehicle CAD model with the roof-mounted UE antenna and its parameters, and Table \ref{tab:table3} lists the C-V2B downlink simulation settings across frequencies.
\begin{table}
\caption{Simulation Parameters for MIMO C\,-V2B Communications}
\label{tab:table3}
\resizebox{\columnwidth}{!}{
\begin{tabular}{|c|c|c|c|c|}
\hline
\textbf{Parameter} & \multicolumn{4}{c|}{\textbf{Value / Description}} \\ \hline
\multicolumn{5}{|c|}{\cellcolor[HTML]{EFEFEF}\textbf{Frequency and Environment}} \\ \hline
Network area (km $\times$ km) & \multicolumn{4}{c|}{$1.2 \times 1.2$ km$^2$} \\ \hline
ISD (m) & \multicolumn{4}{c|}{350 m} \\ \hline
Number of cell sectors & \multicolumn{4}{c|}{3} \\ \hline
Power spectral density of noise (A$^2$/Hz) & \multicolumn{4}{c|}{$10^{-21}$ A$^2$/Hz} \\ \hline
Frequency bands (GHz) & 4.6 GHz & 8.2 GHz & 15 GHz & 28 GHz \\ \hline
Bandwidth (MHz) & 60 MHz & 200 MHz & 300 MHz & 400 MHz \\ \hline
\multicolumn{5}{|c|}{\cellcolor[HTML]{EFEFEF}\textbf{Base Station}} \\ \hline
Location & \multicolumn{4}{c|}{Rooftop} \\ \hline
Antenna element radiation pattern & \multicolumn{4}{c|}{\begin{tabular}[c]{@{}c@{}}HPBW $65^\circ$, maximum element gain 30 dBi\\ (3GPP TR 37.840)\end{tabular}} \\ \hline
Antenna tilt angle & \multicolumn{4}{c|}{$-12^\circ$} \\ \hline
\multirow{2}{*}{URA dimensions} & 4.6 GHz & 8.2 GHz & 15 GHz & 28 GHz \\ \cline{2-5}
& $2\times2$ & $3\times3$ & $5\times5$ & $9\times9$ \\ \hline
\multicolumn{5}{|c|}{\cellcolor[HTML]{EFEFEF}\textbf{User Equipment}} \\ \hline
Location & \multicolumn{4}{c|}{Random locations following a uniform distribution} \\ \hline
Height (m) & \multicolumn{4}{c|}{2 m (3GPP TR 37.840)} \\ \hline
Antenna element radiation pattern & \multicolumn{4}{c|}{Isotropic antenna (3GPP TR 37.840)} \\ \hline
\multirow{2}{*}{ULA dimensions} & 4.6 GHz & 8.2 GHz & 15 GHz & 28 GHz \\ \cline{2-5}
& \multicolumn{2}{c|}{$1\times2$} & \multicolumn{2}{c|}{$1\times3$} \\ \hline
\end{tabular}
}
\end{table}
\begin{table}
\caption{Material properties at different frequencies}
\label{tab:table2}
\centering
\resizebox{\columnwidth}{!}{%
\begin{tabular}{|l|cc|cc|cc|cc|}
\hline
\rowcolor[HTML]{EFEFEF}
\textbf{Material} &
\multicolumn{2}{c|}{\textbf{4.6 GHz}} &
\multicolumn{2}{c|}{\textbf{8.2 GHz}} &
\multicolumn{2}{c|}{\textbf{15 GHz}} &
\multicolumn{2}{c|}{\textbf{28 GHz}} \\ \cline{2-9}
\rowcolor[HTML]{EFEFEF}
& \textbf{$\sigma$ (S/m)} & \textbf{$\varepsilon_r$}
& \textbf{$\sigma$ (S/m)} & \textbf{$\varepsilon_r$}
& \textbf{$\sigma$ (S/m)} & \textbf{$\varepsilon_r$}
& \textbf{$\sigma$ (S/m)} & \textbf{$\varepsilon_r$} \\ \hline
Concrete   & 0.14  & 5.24 & 0.23  & 5.24 & 0.38  & 5.24 & 0.63  & 5.24 \\
Glass      & 0.03 & 6.31 & 0.06 & 6.31 & 0.12 & 6.31 & 0.24 & 6.31 \\
Brick      & 0.03  & 3.91 & 0.03  & 3.91 & 0.04  & 3.91 & 0.04  & 3.91 \\
Dry earth  & 0.003 & 3.00 & 0.01 & 3.00 & 0.036 & 3.00 & 0.147 & 3.00 \\ \hline
\end{tabular}%
}
\end{table}
\begin{figure}[t!]
\centering
\begin{subfigure}[t]{0.48\linewidth}
    \includegraphics[clip,trim=6 1 30 8,scale=0.33]{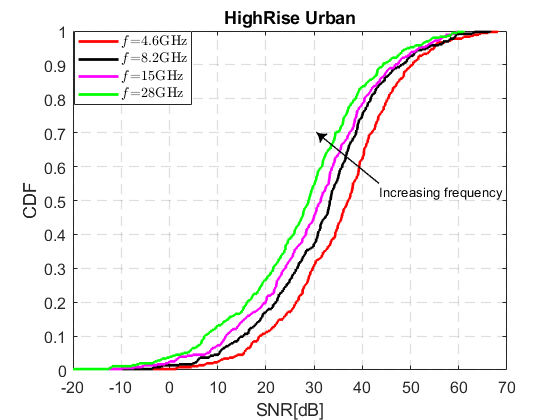}
    \caption{}
\end{subfigure}
\hfill
\begin{subfigure}[t]{0.48\linewidth}
    \includegraphics[clip,trim=6 1 30 8,scale=0.33]{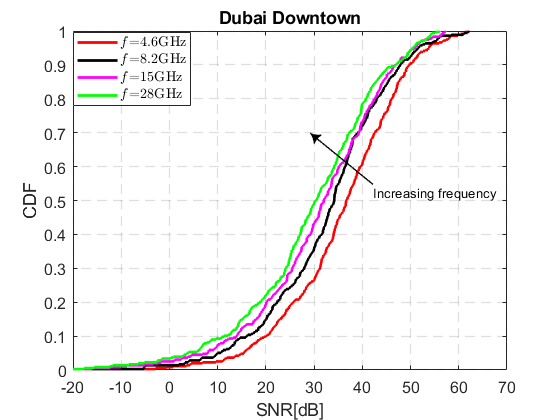}
    \caption{}
\end{subfigure}

\vspace{1mm}

\begin{subfigure}[t]{0.48\linewidth}
    \includegraphics[clip,trim=-7 1 0 8,scale=0.33]{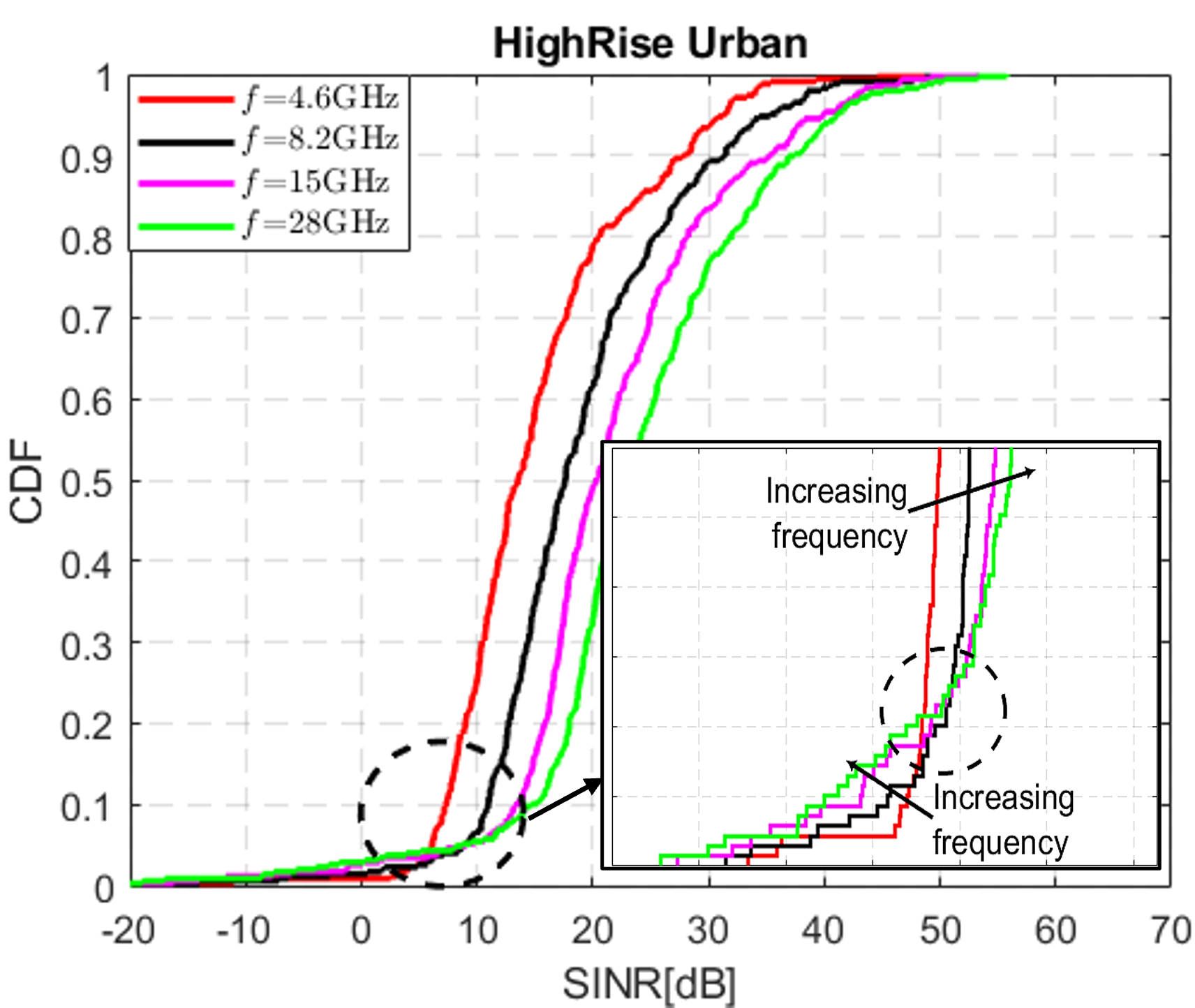}
    \caption{}
\end{subfigure}
\hfill
\begin{subfigure}[t]{0.48\linewidth}
    \includegraphics[clip,trim=-1 1 0 8,scale=0.33]{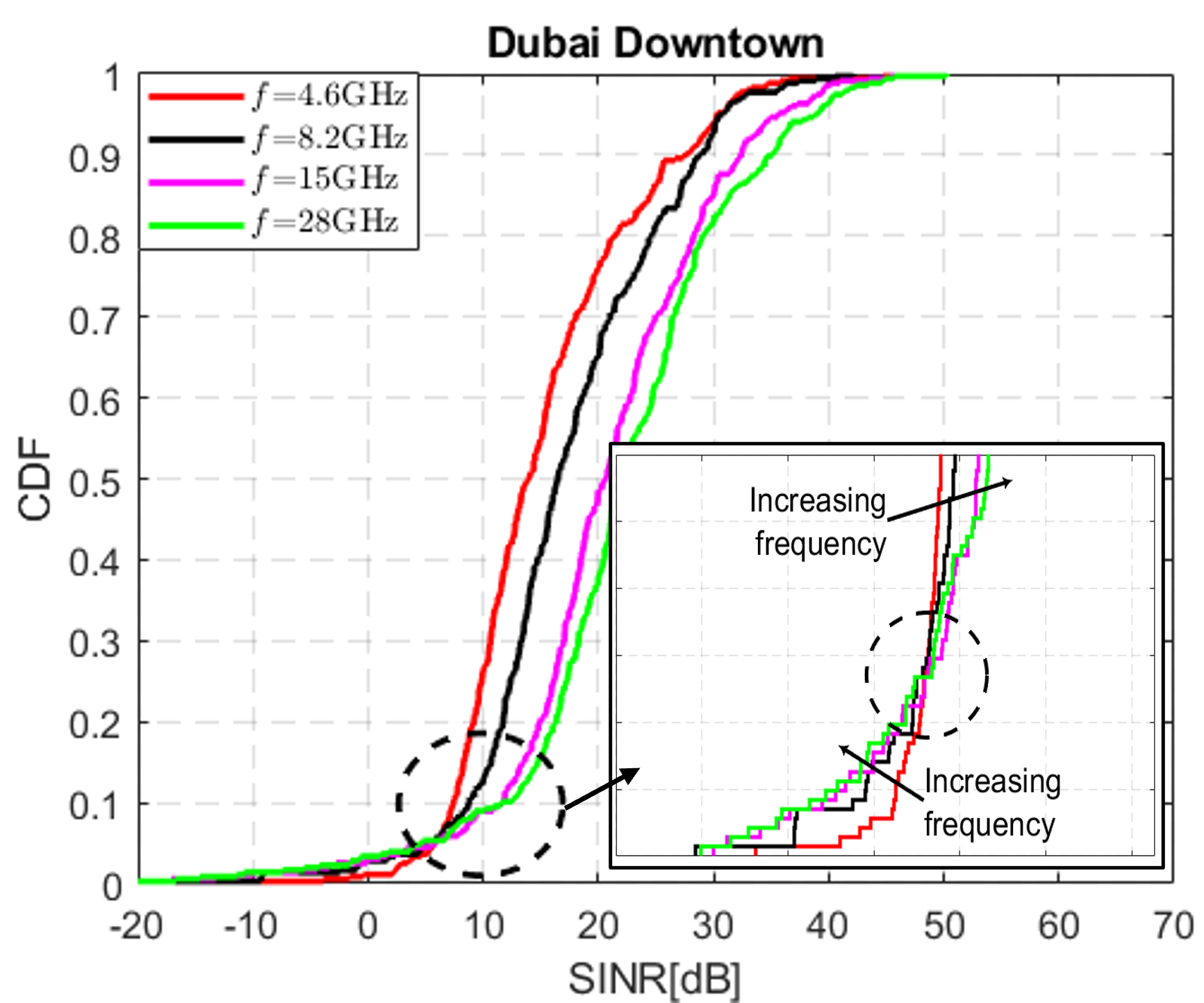}
    \caption{}
\end{subfigure}

\caption{(a–b) SNR CDFs under {\it interference-free} scenario; (c–d) SINR CDFs under {\it full-interference} scenario in HighRise Urban environment by ITU statistical model, and 3D CAD model of the Dubai downtown across different frequencies considering static blockage, respectively.}
\label{fig:SNR}
\end{figure}
\subsection{FR3 vs. Non-FR3 Band Performance}
Fig. \ref{fig:SNR} shows the SNR/SINR CDFs under {\it interference-free} and {\it full-interference} scenarios considering static blockage across the sub-6 GHz, FR3, and mmWave bands at 4.6, 8.2, 15, and 28 GHz, respectively. Figs. \ref{fig:SNR}(a–b) present the SNR CDFs, and Figs. \ref{fig:SNR}(c–d) illustrate the SINR CDFs following the ITU statistical model and a 3D CAD model of the Dubai downtown area, respectively. 
As shown in Fig. \ref{fig:SNR}(a), UEs operating at lower frequencies consistently achieve higher SNRs under Interference-Free cases. In contrast, UEs operating at higher frequencies experience lower SNRs due to the combined influence of several factors, including path loss in numerator of (5), antenna gain, and noise power (i.e., $N_{0}B$) in the denominator which affect all frequency bands. Higher frequencies are more impacted by their larger bandwidths, which increase noise power, and higher path loss. At higher frequencies, larger bandwidths elevate noise power and higher path loss further degrades signal quality. Although narrower beamwidths yield higher antenna gains, this improvement only partially offsets the overall SNR degradation.
Overall, the SNR differences at the 50\% CDF level are approximately 8 dB between the lowest (4.6 GHz) and highest (28 GHz) frequency bands. Compared to the lower sub-6 GHz band (4.6 GHz), FR3 frequencies achieve comparable SNRs at the lower 10\% and upper 90\% CDF levels, benefiting from directional beamforming and moderate path loss. Unlike the mmWave band (28 GHz), UEs operating in the FR3 exhibit higher SNRs across all CDF levels due to the smaller bandwidth, which results in lower noise power. Comparing Fig. \ref{fig:SNR}(a) and \ref{fig:SNR}(b), similar SNR ranges are observed simulated by RT using both the ITU statistical model and the 3D CAD model of Dubai downtown. The two models show approximately 2\%-4\% deviation between their SNR CDFs at higher frequencies attributed to the realistic propagation effects including reflections, diffraction, and partial blockages captured in the 3D model, which mitigate frequency sensitivity. Consequently, the ITU statistical model exhibits stronger frequency dependence with a wider separation between curves, whereas the 3D CAD-based results show closer SNR CDFs, indicating a more uniform channel behavior across frequencies.

To investigate the impact of inter-cell interference in the presence of the static blockage, the {\it full-interference} case is also examined. As a worst-case scenario, all BSs transmit simultaneously to their associated UEs causing each UE to experience interference from neighboring BSs. As shown in Figs. \ref{fig:SNR}(c–d), the SINR CDFs reveal distinct propagation characteristics. Notably, the frequency-order of the curves revers regarding to lower/higher CDF levels (a turning point at $\approx$ 5\% CDF level shown in Fig. \ref{fig:SNR}(c-d)). The CDFs are primarily influenced by static blockage, multipath propagation, antenna gain, and frequency-dependent attenuation.

Three key observations are seen in these figures:
\begin{itemize}
    \item {\bf Observation 1:} {\it In the upper 10\% of the SINR CDF, the curves exhibit a frequency-order inversion compared to the SNR curves}, i.e., {\it higher carrier frequencies yield higher SINR but lower SNR.}
    \\
    {\bf Discussion:} This reversal
occurs because transmission at higher frequency benefits from narrower beamwidths, enabled by fitting more antenna elements within the same transmitter aperture size across all frequencies. The resulting beamforming gain reduces inter-cell interference in the SINR denominator represented in (5)-(6) and offsets the adverse effects of frequency-dependent attenuation as well as the increased noise power associated with wider bandwidths at higher frequencies (See Table II). However, for the SNR expression in (5) under $P_\mathrm{I,avg}$=\,0, the absence of such constructive beamforming gain causes the degradation effects from higher path loss and increased noise power at higher frequencies to dominate. Notably, the SINR gap between 4.6 GHz and 28 GHz at the 50\% CDF level widens to about 10 dB.

    \item {\bf Observation 2:} {\it At the lower 10\% of the CDF level corresponding to UEs located close to the cell edge, both the SNR and SINR degrade with respect to frequency growth (i.e., no performance reversal exist as evidenced in Observation 1.} 
    \\
    {\bf Discussion:} This degradation arises from the sensitivity of both 15 GHz and 28 GHz frequencies to static blockages, as well as larger BS–UE distance typical in HighRise Urban areas, which increase the probability of multiple NLoS links. However, at shorter BS-UE distances, as discussed in {\bf Observation 1}, the higher LoS probability and beamforming gain can partially offset the inter-cell Interference losses while effectively suppressing side-lobe interference. Greater SINR variability and degradation are observed at lower percentiles (up to the $\approx 5\%$ SINR CDF, as shown in Fig. \ref{fig:SNR}(c)). In this region, SINR gaps in the mmWave band range from approximately -25 dB up to 7 dB, reflecting the influence of the path loss and noise power intensifying factors. 

    \item 
    {\bf Observation 3 (FR3 Analysis):}
    {\it
    Both the ITU statistical model and the 3D CAD model of Dubai downtown show that at 
    CDF levels below 10\% and above 90\%, 
     the observed SINR at FR3 is very close to the best SINR corresponding to sub-6 GHz and mmWave bands, respectively.
    }\\
    {\bf Discussion:} 
This trend is driven by the balance between frequency-dependent path loss, beamforming gain determined by the number of array elements, and noise power proportional to bandwidth in FR3, as reflected in SINR comparisons with sub-6 GHz and mmWave bands. Simulations using both the ITU model and the 3D CAD model of Dubai downtown indicate that, at CDF values below 10\%, the SINR results at 8 GHz are close to the best sub-6 GHz performance with a gap approximately 1 dB. Conversely, at CDF values above 90\%, the SINR results at 15 GHz reaches levels comparable to those observed at 28 GHz (differences $\leq $1 dB), benefiting from strong beamforming gain. These results highlight FR3’s potential, showing a negligible performance gap relative to sub-6 GHz and mmWave bands in HighRise Urban areas, where environmental complexity and blockage significantly impact UEs experiencing both low and high signal quality (below 10\% and upper \%90 CDF levels).  
\end{itemize}
\vspace{-0.5em}
\subsection{Impact of BS Density on Coverage Probability}
\begin{figure}[t!]
\centering
\includegraphics[clip,trim=0 0 0 0, scale=0.78]{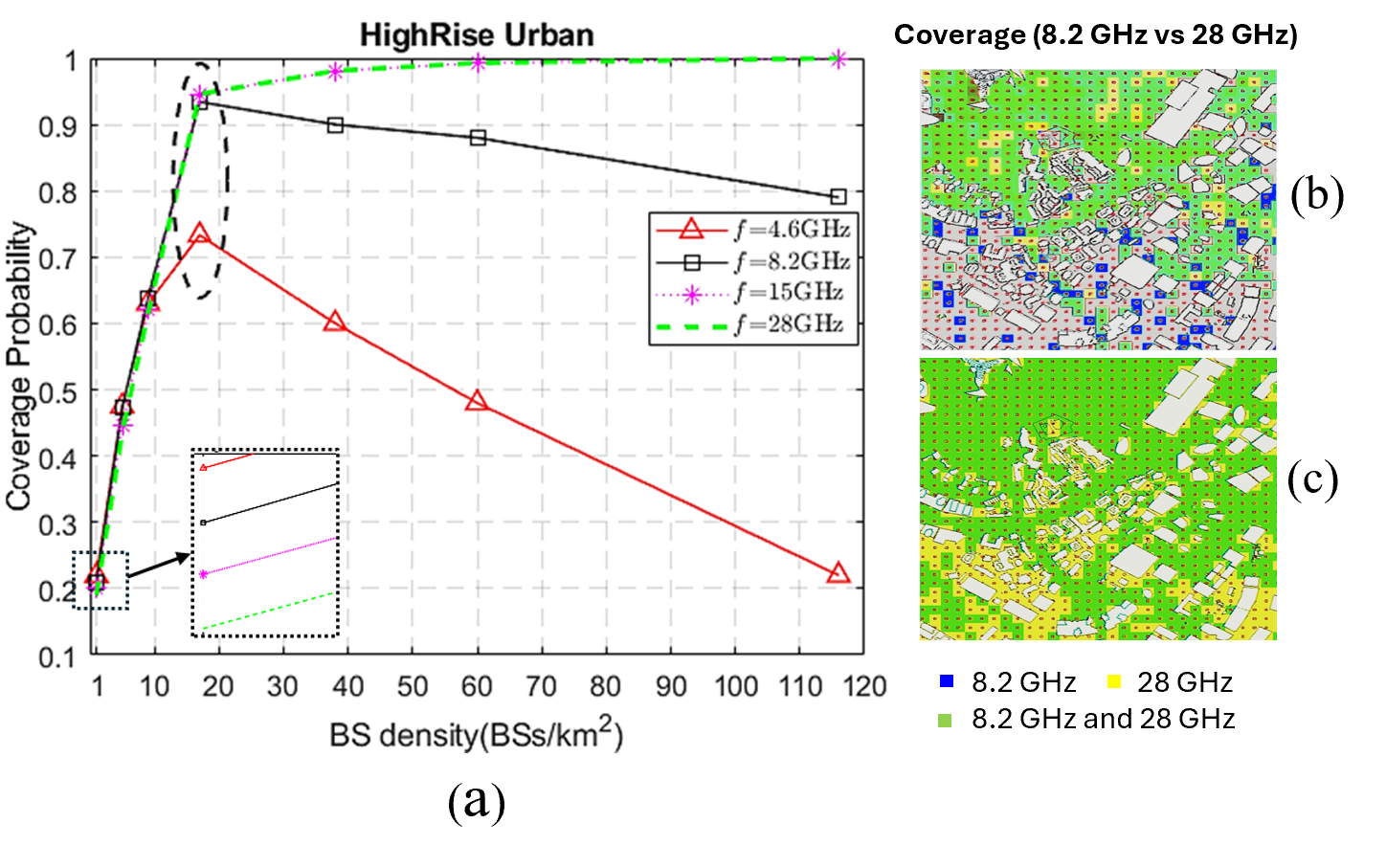}
\caption{\textcolor{black}{Coverage probability and coverage maps, (a)
Coverage probability vs BS density for {\it full-interference} scenario across all frequencies corresponding to $\gamma^{\mathrm{th}}$~= 10 dB, (b) Coverage probability map for BS density of 9 BS/km$^2$, (c) Coverage probability map for BS density of 116 BS/km$^2$.}}
\label{fig:ISD}
 \end{figure}

Based on the statistical evaluation of interference explained in Section IV, the coverage probability is defined as the probability that the SINR of UEs exceeds a given threshold $\gamma^{\mathrm{th}}$, expressed as 
\vspace{-0.5em}
\begin{equation}
p^{\mathrm{cov}}(\gamma^{\mathrm{th}}) = \Pr \left[ \mathrm{SINR}^{\mathrm{UE}} > \gamma^{\mathrm{th}} \right],
\end{equation}
where the SINR is calculated by (5), and $\gamma^{\mathrm{th}}$ is set to 10 dB.
It is seen from Fig. \ref{fig:ISD} that the coverage probability strongly depends on both the operating frequency and the BS deployment density under the {\it full-interference} scenario. Results are presented across all frequencies for BS densities of 1, 5, 9, 17, 38, 60, and 116 BS/km$^{2}$, corresponding to ISDs of approximately 800 m, 650 m, 500 m, 350 m, 200 m, 150 m, and 100 m, respectively. The coverage probability increases with BS densification up to a turning point (highlighted by dashed-line ellipse in Fig. \ref{fig:ISD}(a)) at 17 BS/km$^{2}$ across sub-6 GHz, FR3, and mmWave bands due to enhanced LoS probability and reduced propagation loss. However, for low-frequency bands (e.g., 4.6 GHz and 8.2 GHz), further densification beyond 17 up to 116 BS/km$^{2}$ leads to a gradual degradation in coverage probability driven by increased inter-cell interference. As illustrated in Fig. \ref{fig:ISD}(a), the maximum achievable coverage probability for 4.6 GHz and 8.2 GHz are approximately 0.73 and 0.93, respectively. 
In contrast, systems operating at higher-frequency bands (15 GHz and 28 GHz) experience higher coverage probabilities in dense BS deployments, benefiting from significant beamforming gains of large antenna arrays and the narrower beamwidths, which effectively suppress inter-cell interference. 
The coverage maps in Figs. \ref{fig:ISD}(b) and (c) further validate this trend, showing that at a BS density of 9 BS/km², both 8.2 GHz and 28 GHz achieve nearly equal coverage probabilities ($\approx$ 62\%). However, at BS density of 116 BS/km², the 28 GHz achieves near-unity coverage, while the 8.2 GHz coverage decreases to about 79\%, highlighting the interference-limited behavior of lower-frequency bands in ultra-dense deployments.
\vspace{-0.4em}
\section{CONCLUSION}
We conducted a comprehensive FR3 channel characterization in HighRise Urban networks, comparing it with sub-6 GHz and mmWave bands for C-V2B downlink communication. Using Wireless InSite RT simulations with detailed 3D CAD models and MIMO beamforming, we evaluated SNR and SINR under various interference scenarios. Results show that achieving optimal performance across conditions may require switching between low- and high-frequency bands, increasing cost, licensing requirements, and system complexity. More specifically, for local 6G networks where interference is more likely due to a lack of fully orchestrated spectrum strategies among different operators, we demonstrate that FR3 provides a noticeable and well-balanced advantage over FR1 and FR2, particularly in HighRise Urban scenarios. In contrast, the FR3 band offers a balanced trade-off. Notably, when comparing FR3 and mmWave under {\it equal aperture size}, although mmWave benefits from higher array gain due to more antenna elements, yet, FR3 outperforms mmWave in interference-free SNR across all CDF values and in full-interference SINR for low CDFs (i.e., SINR CDF $<$ 0.1, representing cell-edge UEs). This highlights a key advantage of FR3: {\it the additional array gain of mmWave cannot offset its severe path loss in most communication scenarios}.

\bibliographystyle{IEEEtran}
\bibliography{Rio}
\vspace{12pt}
\end{document}